# Ultrathin Pyrolytic Carbon Films on a Magnetic Substrate

Ahmad Umair [1], Tehseen Z. Raza[2] and Hassan Raza[1,3] *

[1] Department of Electrical and Computer Engineering, University of Iowa, Iowa City, Iowa 52242, USA

[2] Department of Physics and Astronomy, University of Iowa, Iowa City, Iowa 52242, USA

[3] Centre for Fundamental Research, Islamabad, Pakistan

**Abstract**

We report the growth of ultrathin pyrolytic carbon (PyC) films on nickel substrate by using chemical vapor deposition at *1000 °C* under methane ambience. We find that the ultra-fast cooling is crucial for PyC film uniformity by controlling the segregation of carbon on nickel. We characterize the in-plane crystal size of PyC film by using Raman spectroscopy. The Raman peaks at *~1354 $cm^{-1}$* and *~1584 $cm^{-1}$* wavenumbers are used to extract the *D* and *G* bands. The corresponding peak intensities are then used in an excitation energy dependent equation to calculate the in-plane crystal size. Using Raman area mapping, the mean value of in-plane crystal size over an area of *100 μm × 100 μm* is about *22.9 nm* with a standard deviation of about *2.4 nm*.

Keywords: pyrolytic carbon, Raman spectroscopy, layered material.

*Corresponding Author: nstnrg@gmail.com

## 1. Introduction:

Pyrolytic carbon (PyC) is a layered carbon nanomaterial with additional covalent bonding between the graphene layers in the stacking direction. It has high chemical stability, and good electrical and thermal conductivities [1]. Due to its excellent physical and chemical properties, PyC has been used as a thin electrode in metal-insulator-semiconductor capacitors, electrochemical sensing, and various biomedical applications [2–4].

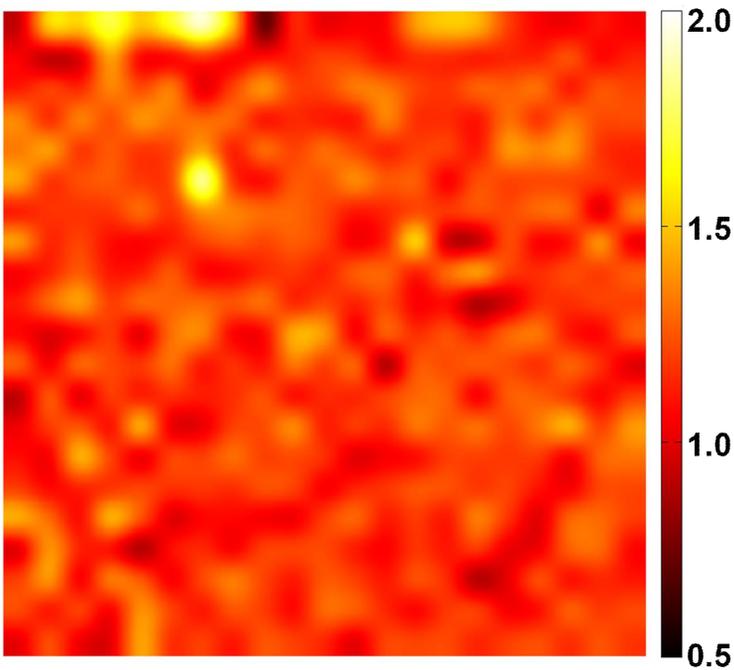

Figure 1 – (color online) Raman Spectroscopy Area Map. Two-dimensional map showing the ratio of the deconvoluted D and G peak intensities from the original spectra. The pixel size is *5 μm × 5 μm* with the scanned area of *100 μm × 100 μm*.

One of the most important material parameter to characterize structural coherence of PyC films is the in-plane crystal size ($L_a$), which can be determined by using the ratio of *D* to *G* peak intensities of the Raman spectrum (shown as a color map in Fig. 1) in a laser excitation energy dependent equation [5,6]. However, additional bands overlap with the *D* and *G* bands in the Raman spectra of PyC films and a deconvolution is imperative to extract the intensities of the *D* and *G* bands.

In this paper, we report the synthesis and characterization of ultrathin PyC films on nickel substrate. We use chemical vapor deposition (CVD) method to synthesize ultrathin PyC films by $CH_4$ decomposition on nickel substrate under ultrafast cooling. Synthesizing PyC films on nickel is advantageous due to the high carbon solubility of carbon in nickel leading to a small growth time. Raman spectroscopy is then used to characterize the film's structural coherence. PyC synthesis on a magnetic substrate like nickel may find applications in spintronics.

This paper is divided into four sections. In Sec. 2, we discuss the CVD growth method. The film characterization is discussed in Sec. 3. Conclusions are provided in Sec. 4.

**2. Method**

PyC films were synthesized on a *300 nm* nickel film evaporated on a *300 nm* $SiO_2$ film, which was thermally grown on a Si substrate. Si/$SiO_2$ substrate was cleaned with *10 minutes* acetone dip, *10 minutes* methanol dip, *10 minutes* deionized (DI) water rinse, *20 minutes* nanostrip (commercial Piranha substitute) dip, followed by another *10 minutes* DI water rinse and nitrogen dry. After cleaning, nickel was evaporated by using electron beam evaporator (Angstrom Engineering) in an alumina crucible. The evaporation rate was *1 Å/s*, and the chamber pressure was *< $10^{-7}$ Torr*. The UV ozone chamber was used to clean the substrate before loading it in the CVD furnace. This helped to eliminate the organic contaminants from the nickel film, which is important for the uniformity of the synthesized PyC film. Research grade (minimum purity *99.999%*) process gases were used. The sample was then loaded into a homemade CVD furnace (Lindbergh/Blue, 1 inch tube diameter) at room temperature and heated to *700 °C* in *200 sccm* (standard cubic centimeter per minute) Ar ambient. At *700 °C*, *65 sccm* $H_2$ was introduced in addition to the Ar and the samples were annealed for *10 minutes*. The temperature was then ramped to *1000 °C* in the Ar:$H_2$ ambient. To stabilize the growth temperature, the samples were annealed for *10* more *minutes* after reaching the *1000 °C* ambient temperature. Finally, $H_2$ was turned off and the PyC was synthesized by introducing *23 sccm* $CH_4$ into the furnace for 40 s, in addition to the already flowing Ar gas. The sample temperature was then suddenly reduced by pulling the quartz tube out of the hot region of the furnace within a few seconds.

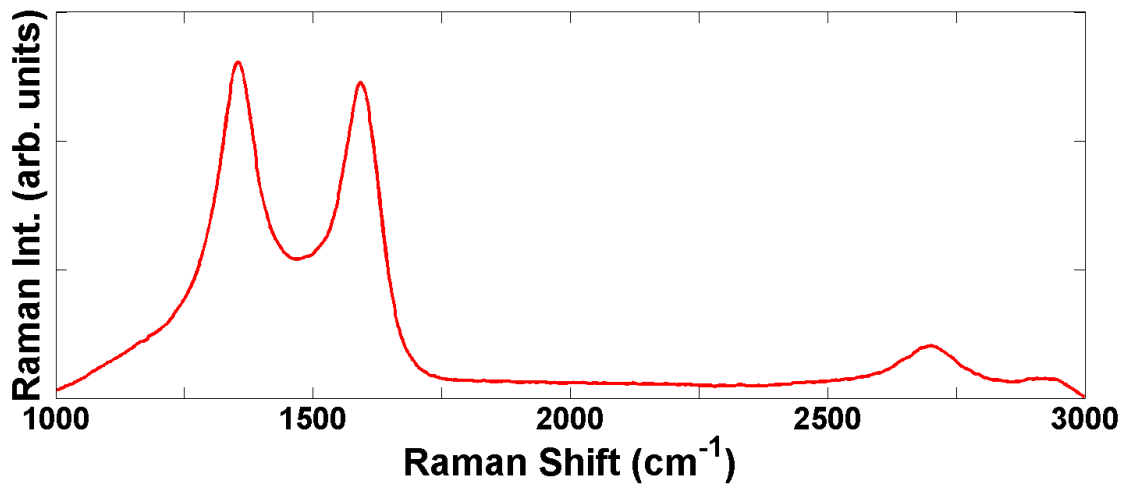

Figure 2 – (color online) Raman spectrum of the PyC film in point scan mode.

We further used Raman spectroscopy (Raman Nicolet Almega XR Spectrometer) in the point scan and the area scan mode to characterize the synthesized PyC films. *532 nm* laser (*10 mW* power) was used in the point scan mode with a *0.6 μm* spot size, *15 seconds* scan time and *4* scans per point. To examine the uniformity of the synthesized PyC, the D and G bands were extracted from the Raman map measured in the area scan mode. The ratio of D to G peak intensities ($I_D/I_G$ ratio) are shown in Fig. 1 over an area of *100 μm × 100 μm*, which is further used in calculating the in-plane crystal size ($L_a$) as we discuss in the next section. For each pixel over the scanned area, a *2.1 μm* spot size was used with *15 seconds* scan time, *4 scans per pixel* and *5 μm* step size.

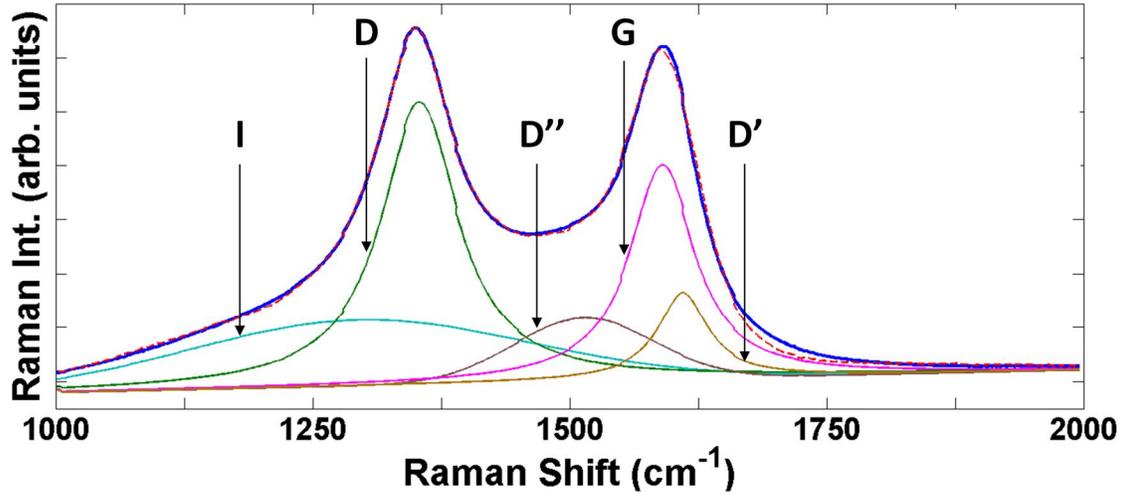

Figure 3 – (color online) Deconvolution of various bands in the low wavenumber range. *D*, *G*, *D'*, *D''* and *I* bands are shown and the corresponding parameters are reported in Table I. Dashed line is the measured data and the solid line is the fit.

Table I – Deconvoluted bands. Fit type, band center and full width half maxima (FWHM) are reported for the deconvoluted bands shown in Fig. 3.

| Band | Fit Type | Band Center (cm$^{-1}$) | FWHM (cm$^{-1}$) |
|---|---|---|---|
| *I* | Gaussian | *1295.00* | *387.33* |
| *D* | Lorentzian | *1353.99* | *89.64* |
| *D''* | Gaussian | *1513.76* | *159.43* |
| *G* | Lorentzian | *1584.00* | *75.52* |
| *D'* | Lorentzian | *1610.00* | *57.85* |

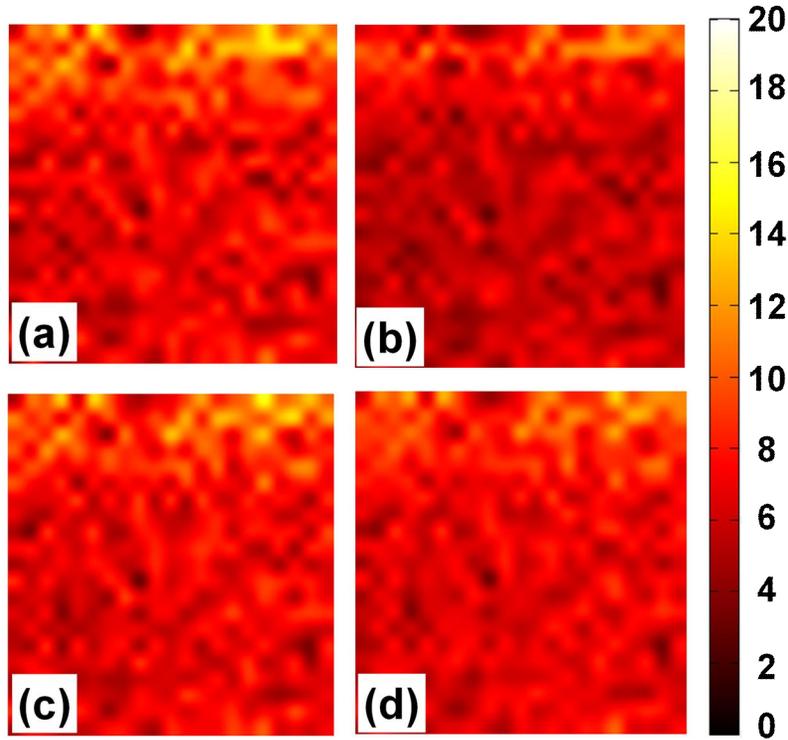

Figure 4 – (color online) Two-dimensional Raman intensity maps. The peaks in the original spectrum are observed at (a) ~$1354\ cm^{-1}$ and (b) ~$1584\ cm^{-1}$ wavenumbers. The intensity maps of the extracted (c) $D$ peaks and (d) $G$ peaks from the deconvoluted spectra. The pixel size is $5\ \mu m \times 5\ \mu m$ with the scanned area of $100\ \mu m \times 100\ \mu m$.

## 3. Results and discussion:

Raman spectrum of the synthesized PyC film in the point scan mode is shown in Fig. 2. The two significant peaks are centered at ~$1354\ cm^{-1}$ and ~$1584\ cm^{-1}$ wavenumbers. These peaks contain $I$, $D'$ and $D''$ bands besides the more commonly observed $D$ and $G$ bands. The $G$ band in the spectrum arises due to the in-plane vibrations of C-C atoms. The $D$ and $D'$ bands are the result of intra- and inter-valley Raman processes due to the defects in the $sp^2$ carbon nanomaterials, respectively [7,8]. The $D''$ and $I$ bands originate from the amorphous carbon and the $sp^2$–$sp^3$ bond combinations, respectively [1]. The Raman spectrum in the low wavenumber range is deconvolved as shown in Fig. 3 to extract the contribution of *I, D, D", G* and *D'* bands. Lorentzian functions are used for fitting *D, G* and *D'* bands, while Gaussian functions are used for *D"* and *I* bands [1]. The extracted parameters for the *I, D, D", G* and *D'* bands are reported in Table I.

The Raman intensity area maps for the peaks at ~*1354 cm$^{-1}$* and ~*1584 cm$^{-1}$* wavenumbers are shown in Figs. 4(a,b), respectively. We deconvolve the Raman spectrum at each scan point and extract the *G* and *D* bands. The area maps of the extracted *D* and *G* peak intensities ($I_D$, $I_G$) are shown in Figs. 4(c,d), respectively; whereas we report the area maps of the ratios of $I_D$ and $I_G$ intensities in Fig. 1. By using this $I_D/I_G$ ratio and the energy or wavelength of the excitation laser, the in-plane crystal size *($L_a$)* for PyC is given as [6-8]:

$$L_a(nm) = \frac{560}{E_{laser}^4}\left[\frac{I_D}{I_G}\right]^{-1} = 2.4 \times 10^{-10} \lambda_{laser}^4 \left[\frac{I_D}{I_G}\right]^{-1}$$

where $E_{laser}$ and $\lambda_{laser}$ are the laser excitation energy (in *eV*) and wavelength (in *nm*), respectively. For a wavelength of *532 nm* used in this study, the pre-factor is *19.2*. In this context, commonly used value of *4.4* as the pre-factor [5] may not be valid in general. Using the above equation, the mean value of the in-plane crystal size over the scanned area is about *22.9 nm*, with a standard deviation of *2.4 nm*. We attribute this uniformity in the in-plane crystal size to the ultra-fast cooling of the samples ó a method that also has also been used for bilayer graphene synthesis [9].

## 4. Conclusions

In summary, we have reported the growth and characterization of uniform ultrathin PyC films on nickel substrate by using CVD and Raman spectroscopy. The segregation of carbon on nickel surface is controlled by the ultra-fast cooling method during CVD. We find that the reported PyC films exhibit small standard deviation in the in-plane crystal size statistics. Although a detailed understanding is yet to be developed, such a cooling method seems to control the segregation of carbon on nickel rather well, enhancing the film's uniformity.

**Acknowledgement:**

We acknowledge the Central Microscopy Facility at the University of Iowa for Raman spectroscopy and the Micro-fabrication facility for metal evaporation. We also thank J. Olesberg, D. Norton, C. Coretsopoulos and J. Baltrusaitis for useful discussions. This work was supported by the University of Iowa, Iowa City USA, and the Associateship program of the Abdus Salam International Center for Theoretical Physics, Trieste Italy.